\documentclass[12pt]{article}
\usepackage{amssymb,amsfonts}
\usepackage{graphics,psboxit,amsmath}
\usepackage{subfigure}
\usepackage{graphicx}
\usepackage{verbatim}
\usepackage{color}
\usepackage{hyperref}
\hypersetup{colorlinks}

\definecolor{darkred}{rgb}{1,0,0}
\definecolor{darkgreen}{rgb}{0,0.5,0}
\definecolor{darkblue}{rgb}{0,0,1}
\definecolor{orange}{rgb}{1,0.5,0}
\definecolor{green}{rgb}{0,1,0}
\definecolor{purple}{rgb}{.5,0,1}

\hypersetup{ colorlinks,
linkcolor=darkred,
filecolor=darkgreen,
urlcolor=darkblue,
citecolor=darkblue,
linktocpage=true }

\definecolor{markcolor}{rgb}{.25,0,1}

\definecolor{markcolor2}{rgb}{1,0,0}

\definecolor{markcolor3}{rgb}{0,1,0}

\def\hybrid{\topmargin -10pt    \oddsidemargin 0.05in 
        \headheight 0pt \headsep 0pt
        \textwidth 16.3cm      
        \textheight 22,2cm       
        \marginparwidth .875in
        \parskip 5pt plus 1pt   \jot = 1.5ex}

\hybrid

\catcode`\@=11

\def\marginnote#1{}
\newcount\hour
\newcount\minute
\newtoks\amorpm
\hour=\time\divide\hour by60 \minute=\time{\multiply\hour by60
\global\advance\minute by-\hour}
\edef\standardtime{{\ifnum\hour<12 \global\amorpm={am}%
        \else\global\amorpm={pm}\advance\hour by-12 \fi
        \ifnum\hour=0 \hour=12 \fi
        \number\hour:\ifnum\minute<10 0\fi\number\minute\the\amorpm}}
\edef\militarytime{\number\hour:\ifnum\minute<10 0\fi\number\minute}

\def\draftlabel#1{{\@bsphack\if@filesw {\let\thepage\relax
   \xdef\@gtempa{\write\@auxout{\string
      \newlabel{#1}{{\@currentlabel}{\thepage}}}}}\@gtempa
   \if@nobreak \ifvmode\nobreak\fi\fi\fi\@esphack}
        \gdef\@eqnlabel{#1}}
\def\@eqnlabel{}
\def\@vacuum{}
\def\draftmarginnote#1{\marginpar{\raggedright\scriptsize\tt#1}}

\def\draft{\oddsidemargin -.5truein
        \def\@oddfoot{\sl preliminary draft \hfil
        \rm\thepage\hfil\sl\today\quad\militarytime}
        \let\@evenfoot\@oddfoot \overfullrule 3pt
        \let\label=\draftlabel
        \let\marginnote=\draftmarginnote
   \def\@eqnnum{(\theequation)\rlap{\kern\marginparsep\tt\@eqnlabel}%
\global\let\@eqnlabel\@vacuum}  }

\def\draft2{
        \def\@oddfoot{\sl preliminary draft \hfil
        \rm\thepage\hfil\sl\today\quad\militarytime}
        \let\@evenfoot\@oddfoot \overfullrule 3pt
        \let\label=\draftlabel
        \let\marginnote=\draftmarginnote
   \def\@eqnnum{(\theequation)\rlap{\kern\marginparsep\tt\@eqnlabel}%
\global\let\@eqnlabel\@vacuum}  }

\def\preprint{\twocolumn\sloppy\flushbottom\parindent 2em
        \leftmargini 2em\leftmarginv .5em\leftmarginvi .5em
        \oddsidemargin -.5in    \evensidemargin -.5in
        \columnsep .4in \footheight 0pt
        \textwidth 10.in        \topmargin  -.4in
        \headheight 12pt \topskip .4in
        \textheight 6.9in \footskip 0pt
        \def\@oddhead{\thepage\hfil\addtocounter{page}{1}\thepage}
        \let\@evenhead\@oddhead \def\@oddfoot{} \def\@evenfoot{} }

\def\numberbysection{\@addtoreset{equation}{section}
        \def\theequation{\thesection.\arabic{equation}}}

\def\underline#1{\relax\ifmmode\@@underline#1\else
        $\@@underline{\hbox{#1}}$\relax\fi}

\def\titlepage{\@restonecolfalse\if@twocolumn\@restonecoltrue\onecolumn
     \else \newpage \fi \thispagestyle{empty}\c@page\z@
        \def\thefootnote{\fnsymbol{footnote}} }

\def\endtitlepage{\if@restonecol\twocolumn \else \newpage \fi
        \def\thefootnote{\arabic{footnote}}
        \setcounter{footnote}{0}}  

\catcode`@=12 \relax

\def\figcap{\section*{Figure Captions\markboth
        {FIGURECAPTIONS}{FIGURECAPTIONS}}\list
        {Figure \arabic{enumi}:\hfill}{\settowidth\labelwidth{Figure
999:}
        \leftmargin\labelwidth
        \advance\leftmargin\labelsep\usecounter{enumi}}}
 \relax
\def\tablecap{\section*{Table Captions\markboth
        {TABLECAPTIONS}{TABLECAPTIONS}}\list
        {Table \arabic{enumi}:\hfill}{\settowidth\labelwidth{Table
999:}
        \leftmargin\labelwidth
        \advance\leftmargin\labelsep\usecounter{enumi}}}
 \relax
\def\reflist{\section*{References\markboth
        {REFLIST}{REFLIST}}\list
        {[\arabic{enumi}]\hfill}{\settowidth\labelwidth{[999]}
        \leftmargin\labelwidth
        \advance\leftmargin\labelsep\usecounter{enumi}}}
 \relax
\makeatletter
\newcounter{pubctr}
\def\publist{\@ifnextchar[{\@publist}{\@@publist}}
\def\@publist[#1]{\list
        {[\arabic{pubctr}]\hfill}{\settowidth\labelwidth{[999]}
        \leftmargin\labelwidth
        \advance\leftmargin\labelsep
        \@nmbrlisttrue\def\@listctr{pubctr}
        \setcounter{pubctr}{#1}\addtocounter{pubctr}{-1}}}
\def\@@publist{\list
        {[\arabic{pubctr}]\hfill}{\settowidth\labelwidth{[999]}
        \leftmargin\labelwidth
        \advance\leftmargin\labelsep
        \@nmbrlisttrue\def\@listctr{pubctr}}}
 \relax
\makeatother

\def\be{\begin{equation}}
\def\ee{\end{equation}}
\def\ba{\begin{eqnarray}}
\def\ea{\end{eqnarray}}

\def\no{\noindent}

\def\IR{\relax{\rm I\kern-.18em R}}

\def\diag{{\rm diag}}

\def\bse{\begin{small}\begin{equation*}}
\def\ese{\end{equation*}\end{small}}

\begin{document}


\renewcommand{\theequation}{\thesection.\arabic{equation}}
\csname @addtoreset\endcsname{equation}{section}

\newcommand{\eqn}[1]{(\ref{#1})}

\begin{titlepage}
\begin{center}
\strut\hfill
\vskip 1.3cm


\vskip .5in

{\Large \bf Jumps and twists in affine Toda field theories}

\vskip 0.5in

{\large \bf Anastasia Doikou}\phantom{x}
\vskip 0.02in
{\footnotesize Department of Mathematics, Heriot-Watt University,\\
EH14 4AS, Edinburgh, United Kingdom}
\\[2mm]
\noindent
{\footnotesize and}

\vskip 0.02in
{\footnotesize Department of Computer Engineering \& Informatics,\\
 University of Patras, GR-Patras 26500, Greece}
\\[2mm]
\noindent
\vskip .1cm


{\footnotesize {\tt E-mail: A.Doikou@hw.ac.uk}}\\

\end{center}

\vskip 1.0in

\centerline{\bf Abstract}

The concept of point-like ``jump'' defects is investigated in the context of affine Toda field theories.
The Hamiltonian formulation is employed for the analysis of the problem. The issue is also addressed when
integrable boundary conditions ruled by the classical twisted Yangian are present. In both periodic and boundary cases
explicit expressions of conserved quantities as well as the relevant Lax pairs and sewing conditions are extracted. It is also observed that in the case of the twisted Yangian the bulk behavior is not affected by the presence of the boundaries.

\no

\vfill

\end{titlepage}
\vfill \eject

\tableofcontents

\section{Introduction}

The presence of impurities in integrable systems has attracted considerable attention over the last years \cite{delmusi}-\cite{doikou-defects}, especially when dealing with physical applications and confronting experimental data. Integrability offers an elegant framework such that impurities may be naturally incorporated to a physical system in a systematic and controllable manner. The algebraic frame describing the presence of point-like defects in both discrete integrable
models and classical integrable field theories is by now well established through the quantum inverse scattering method (QISM) \cite{qism}, and the Hamiltonian formulation respectively \cite{ft}. Both classical and quantum approaches are based on the existence of a defect Lax operator that satisfies the same quadratic quantum algebra as the bulk monodromy matrix.

In the present investigation we implement type-II defects, associated to the classical deformed $\mathfrak{gl}_{n+1}$ algebra, in the context of affine Toda field theories (ATFT's). We examine first the model in the case of periodic or Schwartz boundary conditions, and also consider the defect in the presence of boundary conditions that ruled by the classical twisted Yangian \cite{sklyanin, molev, annecy-twist}. In general, depending on the choice of boundary conditions the bulk physical behavior is accordingly modified. Specifically, in the context of imaginary $A^{(1)}_n$ ATFT the boundary conditions introduced in \cite{corrigan-twist}, known as soliton non-preserving (SNP) are related to the classical twisted Yangian, and force a soliton to reflect to an anti-soliton; these are the boundary conditions we are going to consider here. It is however clear that another possibility exists, i.e. boundary conditions that lead to the reflection of a soliton to itself. These boundary conditions are known as soliton preserving (SP), and have been extensively investigated in the frame of integrable quantum spin chains (see e.g. \cite{spinchain-sp} and references therein). Albeit SP boundary conditions are the obvious ones in the framework of integrable lattice models they remained elusive in the context of ATFT's until their full analysis in \cite{avan-doikou-toda}. On the other hand SNP boundary conditions were investigated through the Bethe ansatz formulation for the first time in \cite{doikou-twist}, whereas higher rank generalizations studied in \cite{annecy-twist}.

The outline of this article is as follows: In the next section we introduce the generic Hamiltonian frame in the presence of point-like impurities. In section 3 we recall some basic notions regarding the $A_n^{(1)}$ ATFT  and also analyze the model in the presence of a local defect. We extract the first charges in involution through the generating function  together with the time components of the Lax pairs and the associated sewing conditions as analyticity conditions imposed on the Lax pair. In section 4 in addition to the defect we also consider non-trivial (SNP) boundary conditions associated to the classical twisted Yangian. In this case only ``half'' of the integrals of motion are conserved (e.g. the Hamiltonian) as expected and the behavior of the defect is essentially intact as was also observed in \cite{nls-twist} in the case of the vector non-linear Schrodinger model. In the last section we discus the results of the present investigation, and we also propose some possible future directions.

\section{Classical integrable models with defect}

The fundamental object used in the analysis that follows is the modified monodromy matrix. The monodromy matrix is modified in a way to include the point-like defect at $x=x_0$ (see e.g. \cite{ft, avan-doikou1} and references therein):
\be
T(L, -L; \lambda) = T^{+}(L, x_0; \lambda)\ {\mathbb L}(x_0; \lambda)\ T^{-}(x_0, -L; \lambda), \label{modT}
\ee
where we define the ``bulk'' monodromy matrices associated to the left and right theories as (see e.g. \cite{ft}):
\be
T^{\pm}(x, y; \lambda) = {\cal P}\exp\{ \int_{y}^x dx'\  {\mathbb U}^{\pm}(x'; \lambda)\}. \label{tpm}
\ee
The latter are solutions of the differential equation
\ba
{\partial T^{\pm}(x,y)\over \partial x} = {\mathbb U}^{\pm}(x,t, \lambda)\ T^{\pm}(x,y). \label{dif0}
\ea
The fact that $T^{\pm}$ is a solution of equation (\ref{dif0}) will be extensively used subsequently for obtaining the relevant integrals of motion.
${\mathbb U}^{\pm}$ is part of the Lax pair ${\mathbb U}^{\pm},\ {\mathbb V}^{\pm}$ that satisfy the linear auxiliary problem.
Let $\Psi$ be a solution of the following set of
equations
\ba
&&{\partial \Psi \over
\partial x} = {\mathbb U}^{\pm}(x,t, \lambda) \Psi \label{dif1}
\\ &&
{\partial  \Psi \over \partial t } = {\mathbb V}^{\pm}(x,t,\lambda) \Psi
\label{dif2}
\ea
${\mathbb U}^{\pm},\ {\mathbb V}^{\pm}$ are in general
$n \times n$ matrices with entries functions of complex valued
fields, their derivatives, and the spectral parameter $\lambda$.
Compatibility conditions of the two differential equations
(\ref{dif1}), (\ref{dif2}) lead to the zero curvature condition
\be
\dot{{\mathbb U}^{\pm}} - {\mathbb V}^{\pm'} + \Big [{\mathbb U}^{\pm},\ {\mathbb V}^{\pm}
\Big ]=0, \label{zecu}
\ee
giving rise to the corresponding
classical equations of motion of the system under consideration. Special care should be taken on the defect point; the zero curvature condition on the point reads as \cite{avan-doikou1, avan-doikou}
\be
{d {\mathbb L}(\lambda) \over d t} = \tilde {\mathbb V}^+(\lambda)\ {\mathbb L}(\lambda) - {\mathbb L}(\lambda)\ \tilde {\mathbb V}^-(\lambda) \label{zcd}
\ee
where $\tilde {\mathbb V}^{\pm}$ are the time components of the Lax pairs on the defect point form the left (right). This is an intricate equation, which arises naturally when studying the continuum limit of discrete integrable theories (see e.g. \cite{avan-doikou1}).
A similar equation involving the derivative with respect to the space coordinate would naturally emerge if one would start the whole analysis considering the time component ${\mathbb V}$ of the Lax pair as the fundamental object (see also some recent relevant results \cite{caudr2}). Equivalently one would end up with such an equation considering the corresponding lattice system in discrete time. This complimentary description would provide an obvious connection with B\"{a}cklund transformations in this context. This is a very interesting issue by itself, which will be fully addressed in a forthcoming publication.

Let us also briefly review the algebraic setting regarding the system in the presence of point like defects, i.e. the Hamiltonian formalism.
The existence of the classical $r$-matrix \cite{sts}, satisfying the classical Yang-Baxter equation
\be
\Big [r_{12}(\lambda_1-\lambda_2),\
r_{13}(\lambda_1)+r_{23}(\lambda_2) \Big]+ \Big
[r_{13}(\lambda_1),\ r_{23}(\lambda_2) \Big] =0,
\ee guarantees
the integrability of the classical system.
The monodromy matrices $T^{\pm}$ as well as the defect matrix ${\mathbb L}$ and consequently the modified monodromy matrix $T(L,-L;\lambda)$ satisfy the classical quadratic algebra \cite{ft}:
\be \Big
\{T^{\pm}_{1}(x,y,t;\lambda_1),\ T^{\pm}_{2}(x,y,t;\lambda_2) \Big \} =
\Big[r_{12}(\lambda_1-\lambda_2),\
T^{\pm}_1(x,y,t;\lambda_1)\ T^{\pm}_2(x,y,t;\lambda_2) \Big ]. \label{basic}
\ee
\be \Big
\{{\mathbb L}_{1}(\lambda_1),\ {\mathbb L}_{2}(\lambda_2) \Big \} =
\Big[r_{12}(\lambda_1-\lambda_2),\
{\mathbb L}_1(\lambda_1)\ {\mathbb L}_2(\lambda_2) \Big ]. \label{rtt2}
\ee
and
\be
\{T_1^+(\lambda_1),\ T_2^-(\lambda_2) \} = 0.
\ee
The latter algebraic relations lead to:
\be
\Big \{tr T(L, -L, ;\lambda_1),\ tr T(L, -L; \lambda_2) \Big \} =0,
\ee
guaranteing the classical integrability of the model.

\section{The $A_2^{(1)}$ ATFT with defect}

We shall focus henceforth on the $A_n^{(1)}$ affine Toda field theories in the presence of point-like defects. In particular, we shall exemplify our investigation using the first non-trivial case that is the $A_2^{(1)}$ model. Similar results were extracted in the context of the sine-Gordon ($A_1^{(1)}$) model in \cite{avan-doikou}. Before we begin with the analysis of the problem at hand it will be useful to introduce some fundamental notions regarding the model.

\subsection{Preliminaries}

First introduce the Lax pairs for the left and right bulk theories consisting of the ${\mathbb U}^{\pm}$ and ${\mathbb V}^{\pm}$ matrices; in the $A_n^{(1)}$ case are given by:
\ba
&& {\mathbb V}^{\pm}(x, t; u) = - {\beta \over 2}\ \partial_x \Phi^{\pm} \cdot H + {m\over 4}\ \Big (u\ e^{{\beta \over 2} \Phi^{\pm} \cdot H}\ E_+\ e^{-{\beta \over 2} \Phi^{\pm} \cdot H} - {1\over u}\  e^{-{\beta \over 2} \Phi^{\pm} \cdot H}\ E_-\ e^{{\beta \over 2} \Phi^{\pm} \cdot H} \Big )\cr
&& {\mathbb U}^{\pm}(x, t; u) = {\beta \over 2}\ \Pi^{\pm} \cdot H + {m\over 4}\ \Big (u\  e^{{\beta \over 2} \Phi^{\pm} \cdot H}\ E_+\ e^{-{\beta \over 2} \Phi^{\pm} \cdot H} + {1\over u}\  e^{-{\beta \over 2} \Phi^{\pm} \cdot H}\ E_-\ e^{{\beta \over 2} \Phi^{\pm} \cdot H}\Big )
\ea
$\Phi^{\pm},\ \Pi^{\pm}$ are $n$ vector fields with components $\phi^{\pm}_i,\ \pi^{\pm}_i$ respectively $i \in \{1, 2, \ldots n\}$ and $u$ is the multiplicative spectral parameter $u=e^{{2\lambda \over n+1}}$.
It is useful to note that the Lax pair obeys the following symmetries:
\be
{\mathbb V}^{\pm t}(x, t;-u^{-1}) = {\mathbb V}^{\pm}(x,t;u), ~~~~~{\mathbb U}^{\pm t}(x,t;u^{-1}) = {\mathbb U}^{\pm}(x,t; u),
\ee
where $^t$ denotes usual transposition. The associated classical $r$-matrix is:
\be
r(\lambda) = {\cosh(\lambda) \over \sinh(\lambda)} \sum_{i=1}^{n+1}e_{ii} \otimes e_{ii} +{1\over \sinh(\lambda)}\sum_{i\neq j=1}^{n+1} e^{[sgn(i-j)-(i-j){2\over n+1}\lambda]} e_{ij}\otimes e_{ji}
\ee
where we define the $(n+1)\times (n+1)$ matrices with entries: $(e_{ij})_{kl} = \delta_{ik}\ \delta_{jl}$.

We also define:
\be
E_+ = \sum_{i=1}^{n+1} E_{\alpha_i}
\ee
$\alpha_i$ are the simple roots, $H$ and $E_{\alpha_i}$ the algebra generators in the Cartan-Weyl basis
corresponding to simple roots; they satisfy the following Lie algebra relations:
\ba
&& \Big [H,\ E_{\alpha_i} \Big ] = \pm \alpha_i E_{\alpha_i} \cr
&& \Big [E_{\alpha_i},\ E_{-\alpha_i}\Big ] = {2 \over \alpha_i^2}\ \alpha_i \cdot H.
\ea
The vectors $\alpha_i = \Big (\alpha_i^{1}, \ldots, \alpha_i^n \Big )$ are the simple roots of the rank $n$ Lie algebra
normalized to unity $\alpha_i \cdot \alpha_i = 1$ (see e.g. \cite{georgi}):
\be
\alpha_i = \Big (0, 0, \ldots, - \sqrt{i-1\over 2i},\ \sqrt{i+1 \over 2i}, 0, \ldots , 0 \Big ), ~~~~i \in \{1, \ldots ,n \}.
\ee
The fundamental weights are also defined as: $\mu_k =\Big ( \mu_k^1, \ldots, \mu_k^n\Big )$, $~k\in \{1, \ldots, n\}$ and
\be
\alpha_j\cdot\mu_k = {1\over 2} \delta_{jk}.
\ee
The affine root $\alpha_{n+1}$ is derived by the relation
\be
\sum_{j=1}^{n+1} \alpha_j =0.
\ee
We provide below the Cartan-Weyl generators in the fundamental representation
\ba
&& E_{\alpha_i} = e_{i\ i+1}, ~~~~E_{-\alpha_i} = e_{i+1\ i},~~~~ E_{\alpha_n} = - e_{n+1\ n}, ~~~~ E_{-\alpha_n} = - e_{n\ n+1}, \cr
&& H_i = \sum_{j=1}\mu_j^i(e_{jj} - e_{j+1 j+1}).
\ea

Let us now focus on the $A_2^{(1)}$ case, which is our main interest here. The simple roots in this case are:
\be
\alpha_1 = (1,\ 0), ~~~~\alpha_2 = (-{1\over 2},\ {\sqrt{3} \over 2} ),~~~~~\alpha_3 = (-{1\over 2},\ -{\sqrt{3} \over 2} )
\ee
The corresponding Cartan-Weyl generators are given as $3 \times 3$ matrices
\ba
&& E_{1} = E_{-1}^t = e_{12}, ~~~~E_{2} = E_{-2}^t = e_{23}, ~~~~~E_3 = E_{-3}^t = -e_{31} \cr
&& H_1 = {1\over 2} (e_{11} - e_{22}),~~~~~H_2 = {1\over 2 \sqrt{3}}(e_{11} + e_{22} - 2 e_{33}).
\ea

Our first aim is to derive the relevant local integrals of motion via the expansion of the generating function.
To achieve this we are going to exploit the following relations:
\ba
&& T^{\pm}(x,y;\lambda) = \Omega^{\pm}(x)\ \tilde T^{\pm}(x,y;\lambda)\ (\Omega^{\pm}(y))^{-1}, \cr
&& T^{\pm}(x,y; \lambda) = (\Omega^{\pm}(x))^{-1}\ \hat{\tilde T}^{\pm}(x,y;\lambda)\  \Omega^{\pm}(y) \label{dec}
\ea
where
\be
\Omega^{\pm} = e^{{\beta \over 2} \Phi^{\pm} \cdot H}= \diag \Big (\alpha^{\pm},\ \beta^{\pm}, \gamma^{\pm}\Big ), \label{deff1}
\ee
as well as the standard decomposition ansatz for the $\tilde T,\ \hat{\tilde T}$ matrices
\ba
&& \tilde T^{\pm}(x, y; u) =(1 +W^{\pm}(x,u))\ e^{Z^{\pm}(x,y;u)}\ (1+W^{\pm}(y,u))^{-1} \cr
&& \hat {\tilde T}^{\pm}(x, y; u) =(1 +\hat W^{\pm}(x,u))\ e^{\hat Z^{\pm}(x,y;u)}\ (1+\hat W^{\pm}(y,u))^{-1} \label{dec2}
\ea
$Z^{\pm},\ \hat Z^{\pm}$ and $W^{\pm},\ \hat W^{\pm}$ are diagonal and off diagonal $(n+1) \times (n+1)$ matrices respectively such that:
\ba
&& W^{\pm}(u) = \sum_{k=0}^{\infty} {W^{\pm(k)} \over u^k}, ~~~~~Z^{\pm}(u) = \sum_{k=-1}^{\infty} {Z^{(k)} \over u^k},\cr
&& \hat W^{\pm}(u) = \sum_{k=0}^{\infty} {\hat W^{\pm(k)} \over u^{-k}}, ~~~~~ \hat Z^{\pm}(u) = \sum_{k=-1}^{\infty} {\hat Z^{\pm(k)} \over u^{-k}}. \label{wz}
\ea
Substituting the ansatz (\ref{dec}), (\ref{dec2}) in (\ref{dif1}), and considering the diagonal and off diagonal contributions separately we may identify the $W^{\pm}$ and $Z^{\pm}$ $(\hat Z^{\pm},\ \hat W^{\pm})$ matrices (see the Appendix for more details on the identification of $W^{\pm}$ and $Z^{\pm}$ $(\hat Z^{\pm},\ \hat W^{\pm})$, see also \cite{avan-doikou-toda}).

\subsection{The model with defect}

We come now to our main aim, which is the derivation of the charges in involution as well as the identification of the time component of the Lax pair. We choose to consider here the following type-II defect ${\mathbb L}$-operator for the $A_{2}^{(1)}$ model
\ba
{\mathbb L}(\lambda) = \begin{pmatrix}
e^{\lambda}e^{\varepsilon_1} -e^{-\lambda} e^{-\varepsilon_1}& e^{-{\lambda\over 3}}\ t_{12} & e^{\lambda \over 3}\ t_{13}\\
e^{\lambda\over 3}\ t_{21} &e^{\lambda}e^{\varepsilon_2} -e^{-\lambda} e^{-\varepsilon_2}  & e^{-{\lambda \over 3}}\ t_{23}\\
e^{-{\lambda \over 3}}\ t_{31} & e^{\lambda\over 3}\ t_{32}   &e^{\lambda}e^{\varepsilon_3} -e^{-\lambda} e^{-\varepsilon_3}
\end{pmatrix}.
\ea
$t_{ij},\ \varepsilon_i$ are naturally dynamical objects, with Poisson commuting relations dictated by (\ref{rtt2}). In fact, they form a deformed version of the classical $\mathfrak{sl}_3$ algebra.
Note that the defect matrix as well as the classical $r$-matrix of the model are expressed in the so-called {\it principal gradation}.

The generating function of the local integrals of motion is given as
\be
{\cal G}(u) = \ln t(\lambda) = \ln tr \Big ( T^+(L,x_0; u)\ {\mathbb L}(x_0; u)\   T^-(x_0, -L;u)\Big ).
\ee
Expansion of the latter expression in powers of $u^{-1},\ u $ yields the associated integrals of motion as will be clear below. We implement the decomposition ansatz (\ref{dec}), (\ref{dec2}) and expand in powers of $u,\ u^{-1}$.
Then the following expressions arise:
\ba
{\cal G}(u) &=& Z^+_{33}(L, x_0;u) + Z^{-}_{33}(x_0, -L; u) + \ln X(x_0;u)
\cr
{\cal G}(u) &=& \hat Z^+_{33}(L, x_0; u) + \hat  Z^{-}_{33}(x_0, -L; u) + \ln \hat X(x_0;u)
\ea
where we define:
\ba
&& X(x_0; u) = \ln \Big [ (1+W^+(x_0;u))^{-1} (\Omega^+(x_0))^{-1} {\mathbb L}(x_0;u)\Omega^-(x_0) (1+W^-(x_0;u))\Big ]_{33} \cr
&& \hat X(x_0; u)= \ln \Big [ (1+\hat W^+(x_0;u))^{-1} \Omega^+(x_0){\mathbb L}(x_0; u) (\Omega^-(x_0))^{-1} (1+\hat W^-(x_0;u))\Big ]_{33}.
\ea
We also assume here Schwartz type boundary conditions at $x= L,\ - L$. Let us at this point introduce some useful notation:
\ba
&& T_{13} = (\alpha^+)^{-1} \gamma^- t_{13}, ~~~T_{21}= (\beta^+)^{-1}\alpha^- t_{21}, ~~~T_{32} = (\gamma^+)^{-1} \beta^- t_{32} \cr
&& \hat T_{12} = \alpha^+(\beta^-)^{-1} t_{12}, ~~~\hat T_{23} = \beta^+ (\gamma^-)^{-1} t_{23},
~~~~\hat T_{31} = \gamma^+ (\alpha^-)^{-1}t_{31} \cr
&& E_1 = e^{\varepsilon_1}(\alpha^+)^{-1} \alpha^-, ~~~~~E_2 =  e^{\varepsilon_2}(\beta^+)^{-1} \beta^-, ~~~~~E_3 =  e^{\varepsilon_3} (\gamma^+)^{-1} \alpha^-, ~~~~~\tilde E_i = - E_i^{-1} \cr &&
\ea
$\alpha^{\pm},\ \beta^{\pm}, \gamma^{\pm}$ are defined in (\ref{deff1}).

Expansion of the quantities $X,\ \hat X$ then yields:
\ba
&& X(u) = X^{(0)} + u^{-1}\ X^{(1)} + \dots, \cr
&& \hat X(u) = \hat X^{(0)} + u\ \hat X^{(1)} + \ldots
\ea
where we define:
\ba
&& X^{(0)} = {1\over 3}\sum_{i=1}^3 E_i , ~~~~\hat X^{(0)}= {1\over 3} \sum_{i=1}^3 \tilde E_i \cr
&& X^{(1)} = {4\over 3 m} \Big (E_1 (c^- +b^+) -E_2 b^- -E_3 c^+ \Big ) + {1\over 3} (T_{13} - T_{21} - T_{32})\cr
&& \hat X^{(1)} = {4\over 3m} \Big ( -\tilde E_1 \hat a^- + \tilde E_2 (\hat c^- + \hat a^+) -\tilde E_3 \hat c^+ \Big )
+ {1\over 3} (\hat T_{31} - \hat T_{12} -\hat T_{23}).
\ea
the quantities $a^{\pm},\ b^{\pm},\ c^{\pm},\ \hat a^{\pm},\ \hat b^{\pm},\ \hat c^{\pm} $ are defined in the Appendix.
Then the generating function may be expressed as:
\ba
&& {\cal G}(\lambda) = Z_{33}^+(u) + Z_{33}^-(u) + \ln X^{(0)} + u^{-1}(X^{(0)})^{-1} X^{(1)} + \ldots \cr
&& {\cal G}(\lambda) = \hat Z_{33}^+(u) + \hat Z_{33}^-(u) + \ln \hat X^{(0)} + u(\hat X^{(0)})^{-1} \hat X^{(1)} + \ldots
\ea
The first two integrals of motion are the momentum and the Hamiltonian of the system defined as:
\ba
&& H  = -{6m \over \beta^2}\Big ( Z_{33}^{+(1)} + Z_{33}^{-(1)} + \hat Z_{33}^{+(1)} + \hat Z_{33}^{-(1)} + X^{(1)} (X^{(0)})^{-1} +  \hat X^{(1)} (\hat X^{(0)})^{-1} \Big ) \cr
&& P = {6m \over \beta^2}\Big ( Z_{33}^{+(1)} + Z_{33}^{-(1)} - \hat Z_{33}^{+(1)} - \hat Z_{33}^{-(1)} + X^{(1)} (X^{(0)})^{-1}- \hat X^{(1)} (\hat X^{(0)})^{-1}\Big ).
\ea
It is expected that the zero order terms should be trivially equal to the unit, that is $E_i = -\tilde E_i = 1$; this will be more transparent in the subsequent section when requiring analyticity conditions on the time components of the Lax pairs. The explicit expressions of the latter quantities ($H,\ P$) are then given by
\ba
&& H = \int_{-L}^{x_0^-} dx \sum_{i=1}^2 \Big( (\pi_i^-)^2+(\phi_i^{-'})^{2} + {m^2 \over \beta^2} \sum_{i=1}^3 e^{\beta \alpha_i \cdot \Phi^-}\Big) \cr &&+ \int_{x_0^+}^{L} dx \sum_{i=1}^2 \Big( (\pi_i^+)^2+(\phi_i^{+'})^{2} +{m^2 \over \beta^2} \sum_{i=1}^3 e^{\beta \alpha_i \cdot \Phi^+}\Big)\cr
&& -{6m \over \beta^2} \Big ({4 \over 3m} (c^- + b^+ -b^- - c^+-\hat \alpha^- + \hat c^-+\hat \alpha^+- \hat c^+) + {1 \over 3} (T_{13} - T_{21} - T_{32}-\hat T_{31} + \hat T_{12} + \hat T_{23})\Big )\Big |_{x=x_0} \cr
&& P = \int_{-L}^{x_0^-} dx  \sum_{i=1}^2 \Big (\pi_i^- \phi_i^{-'} - \pi_i^{-'} \phi_i^{-} \Big ) +  \int_{x_0^+}^{L} dx  \sum_{i=1}^2 \Big ( \pi_i^+ \phi_i^{+'} - \pi_i^{+'} \phi_i^{+} \Big ) \cr
&& {6m \over \beta^2} \Big ({4 \over 3m} (c^- + b^+ -b^- - c^+ + \hat \alpha^- - \hat c^- - \hat \alpha^++ \hat c^+) + {1 \over 3} (T_{13} - T_{21} - T_{32} + \hat T_{31} - \hat T_{12} - \hat T_{23})\Big )\Big |_{x=x_0} \cr
&&
\ea

\subsection*{The Lax pair}

Having determined the conserved quantities of the model in the presence of the point like defect we are ready to extract the associated time components of the Lax pair. It is worth noting that the derivation of the associated Lax pair is somehow necessary in order to extract the sewing conditions across the defect point (see e.g. \cite{avan-doikou}). It was explicitly computed in \cite{avan-doikou} that the corresponding quantities are given as
\ba
&& {\mathbb V}^{\pm}(x;\lambda, \mu) = t^{-1}(\lambda)\ tr_a \Big (M_{ab}^{\pm}(x; \lambda, \mu)  \Big ) \cr
&& \tilde {\mathbb V}^{\pm}(x_0; \lambda, \mu)=  t^{-1}(\lambda)\ tr_a \Big ( \tilde M_{ab}^{\pm}(x_0;\lambda, \mu) \Big )
\ea
The quantities ${\mathbb V}^{\pm}$ are associated to the right and left bulk theories, whereas $\tilde {\mathbb V}^{\pm}$ are associated to the Lax pair on the defect point from the left and right respectively. We also define
\ba
&& M_{ab}^+(x; \lambda, \mu) = T_a^+(L, x;\lambda)\ r_{ab}(\lambda -\mu)\ T_a^+(x, x_0;\lambda)\ {\mathbb L}(x_0; \lambda)\ T^-_a(x_0, -L;\lambda), ~~~x>x_0 \cr
&& M_{ab}^-(x; \lambda, \mu) = T_a^+(L,x_0;\lambda)\ {\mathbb L}(x_0;\lambda)\ T_a^-(x_0,x;\lambda)\ r_{ab}(\lambda -\mu)\ T_a^-(x, -L;\lambda), ~~~x<x_0 \cr
&& \tilde M_{ab}^+ = T_a^+(L,x_0;\lambda)\ r_{ab}(\lambda -\mu)\ {\mathbb L}(x_0;\lambda)\ T_a^-(x_0, -L;\lambda) \cr
&& \tilde M_{ab}^- = T_a^+(L,x_0;\lambda)\ {\mathbb L}(x_0;\lambda)\ r_{ab}(\lambda -\mu)\ T_a^-(x_0, -L;\lambda) \label{mm}
\ea

Implementing the decomposition ansatz (\ref{dec}), (\ref{dec2}) for the monodromy matrices, and expanding in powers of $u^{-1},\ u$ we get the the various orders. Due to analyticity conditions on the zero order terms
\be
\tilde {\mathbb V}^{\pm (0)}(x_0) \to {\mathbb V}^{\pm (0)}(x_0^{\pm}), ~~~~\hat {\tilde {\mathbb V}}^{\pm (0)}(x_0) \to \hat {\mathbb V}^{\pm (0)}(x_0^{\pm})
\ee
we conclude that $\tilde {\mathbb V}^{\pm(0)}, \hat {\tilde {\mathbb V}}^{\pm(0)} \propto {\mathbb I}$, which leads to: $E_i = - \tilde E_i =1$, as already pointed out earlier in the text when deriving the integrals of motion.

Also obtain the following first order terms on the defect point ($v = e^{{2\mu \over 3}}$):
\ba
&& \tilde {\mathbb V}^{-(1)}(x_0; v) = {2v \over 3}\Big ( (\alpha^{-})^{-1}\gamma^- e_{31} - (\beta^-)^{-1} \alpha^- e_{12} - (\gamma^-)^{-1} \beta^- e_{23} \Big )\cr
&& + {1\over 9} \Big ( (-2 T_{21} -T_{13} + T_{32})e_{11}+ (-2 T_{32} - T_{13} +T_{21})e_{22} + (2 T_{13} + T_{21}+ T_{32})e_{33} \Big ) \cr
&&+ {4 \over 9 m} \Big ( ( - a^- -a^+ + c^- +  b^+ )e_{11}+ (-b^- -b^+ + c^+ + a^-)e_{22}+(-c^+ - c^- + a^+ + b^-)e_{33} \ \Big ) \cr
&& \hat {\tilde {\mathbb V}}^{-(1)}(x_0; v) = -{2v^{-1} \over 3}\Big ( -\alpha^{-}(\beta^-)^{-1} e_{21} - \beta^- (\gamma^-)^{-1} e_{32} + \gamma^- (\alpha^-)^{-1} e_{13} \Big )\cr
&& + {1\over 9} \Big ( (2 \hat T_{31} + \hat  T_{12} + \hat T_{23})e_{11}+ (-2 \hat T_{12} - \hat T_{31} +\hat T_{23})e_{22} + (-2 \hat T_{23} + \hat T_{12}- \hat T_{31})e_{33} \Big ) \cr
&&- {4 \over 9m} \Big ( ( - \hat a^- -\hat a^+ + \hat c^+ +  \hat b^- )e_{11}+ (-\hat b^- -\hat b^+ + \hat c^- + \hat a^+)e_{22}+(-\hat c^+ - \hat  c^- + \hat a^- + \hat b^+)e_{33} \ \Big ) \cr
&&
\ea
and the $\tilde {\mathbb V}^{\pm}$ quantities:
\ba
&& \tilde {\mathbb V}^{+(1)}(x_0; v) = {2v \over 3}\Big ( (\alpha^{+})^{-1}\gamma^+ e_{31} - (\beta^+)^{-1} \alpha^+ e_{12} - (\gamma^+)^{-1} \beta^+ e_{23} \Big )\cr
&& + {1\over 9} \Big ( (2 T_{13} +T_{21} + T_{32})e_{11}+ (-2 T_{21} - T_{13} +T_{32})e_{22} + (-2 T_{32} - T_{13}+ T_{21})e_{33} \Big ) \cr
&&+ {4 \over 9m} \Big ( ( - a^+ -a^- + c^- +  b^+ )e_{11}+ (-b^- -b^+ + c^+ + a^-)e_{22}+(-c^+ - c^- + a^+ + b^-)e_{33} \ \Big ) \cr
&& \hat {\tilde {\mathbb V}}^{+(1)}(x_0; v) = -{2v^{-1} \over 3}\Big ( -\alpha^+(\beta^+)^{-1} e_{21} - \beta^+ (\gamma^+)^{-1} e_{32} + \gamma^+ (\alpha^+)^{-1} e_{13} \Big )\cr
&& + {1\over 9} \Big ( (-2 \hat T_{12} + \hat  T_{23} - \hat T_{31})e_{11}+ (-2 \hat T_{23} + \hat T_{12} -\hat T_{31})e_{22} + (2 \hat T_{31} + \hat T_{12}+ \hat T_{23})e_{33} \Big ) \cr
&&- {4 \over 9m} \Big ( ( - \hat a^- -\hat a^+ + \hat c^+ +  \hat b^- )e_{11}+ (-\hat b^- -\hat b^+ + \hat c^- + \hat a^+)e_{22}+(-\hat c^+ - \hat  c^- + \hat a^- + \hat b^+)e_{33} \ \Big ). \cr
&&
\ea

The bulk parts associated to the left and right bulk theories are given by the familiar expressions:
\ba
&& {\mathbb V}^{\pm(1)}(x;v) = -{4\over 3m} \Big (a^{\pm} e_{11} + b^{\pm} e_{22} + c^{\pm} e _{33} \Big ) +{2v \over 3}\Big ((\alpha^{\pm})^{-1} \gamma\ e_{31} - \beta^{\pm} (\alpha^{\pm})^{-1} e_{12} - (\gamma^{\pm})^{-1} \beta^{\pm} e_{23} \Big ) \cr
&& \hat {{\mathbb V}}^{\pm(1)}(x;v) = {4\over 3m} (\hat a^{\pm} e_{11} + \hat  b^{\pm} e_{22} + \hat c^{\pm} e_{33}) -{2v^{-1} \over 3}  \Big (-\alpha^{\pm}(\beta^{\pm})^{-1} e_{21} - \beta^{\pm} (\gamma^{\pm})^{-1} e_{32} + \gamma^{\pm} (\alpha^{\pm})^{-1} e_{13} \Big ) \cr
&&
\ea
The time components corresponding to the Hamiltonian and momentum are given as
\ba
&& {\mathbb V}_{H}^{\pm} \propto {\mathbb V}^{\pm(1)} + \hat  {\mathbb V}^{\pm(1)}, ~~~{\mathbb V}_{P}^{\pm} \propto {\mathbb V}^{\pm(1)} - \hat  {\mathbb V}^{\pm(1)}\cr
&& \tilde {\mathbb V}_{H}^{\pm} \propto \tilde {\mathbb V}^{\pm(1)} + \tilde {\hat  {\mathbb V}}^{\pm(1)}, ~~~\tilde {\mathbb V}_{P}^{\pm} \propto \tilde {\mathbb V}^{\pm(1)} -  \tilde  {\hat {\mathbb V}}^{\pm(1)}. \label{vv}
\ea
Explicit sewing conditions may be derived as analyticity condition on the Lax pairs
\be
\tilde  {\mathbb V}_{H, P}^{\pm}(x_0)\ \to \ {\mathbb V}^{\pm}_{H,P}(x_0^{\pm})
\ee
and have the following form:
\ba
&& \phi_1^{+'} - \phi_1^{-'}= {m\over 4 \beta} \Big ( 2T_{21} -T_{32} +T_{13} -2\hat T_{12} + \hat T_{23} -\hat T_{31} \Big ) \cr
&& \phi_2^{+'} - \phi_2^{-'} = {\sqrt{3} m \over 4 \beta} \Big (T_{32} + T_{13} -\hat T_{23} - \hat T_{31} \Big ) \cr
&& \pi_1^{+} - \pi_1^{-} ={m\over 4 \beta} \Big ( -2T_{21} +T_{32} -T_{13} -2\hat T_{12} + \hat T_{23} -\hat T_{31} \Big ) \cr
&& \pi_2^{+} - \pi_2^{-} = {\sqrt{3} m \over 4 \beta} \Big (-T_{32} - T_{13} -\hat T_{23} - \hat T_{31} \Big ).
\ea
The latter conditions are invariant under the Hamiltonian action as is also discussed in \cite{avan-doikou}. We refer the interested reader to \cite{avan-doikou} for a more detailed discussion on the consistency of the approach on integrable defects.
Note that similar analyticity conditions were derived in \cite{avan-doikou} in the context of the $A_1^{(1)}$ (sine-Gordon) model.

\section{Twisted Yangian and defects in the $A_2^{(1)}$ ATFT}

We shall focus henceforth on the case where in addition to the presence of the point-like defect non-trivial boundary conditions are also considered. We shall be in particular interested in the case where boundary conditions ruled by the classical twisted Yangian are implemented (see also \cite{nls-twist}, and references therein for detailed description). The boundary conditions associated to the reflection algebra present special interest as well as technical challenges, and display highly unusual bulk behavior, therefore will be left for separate investigations.

The classical twisted Yangian \cite{sklyanin} describes the soliton non-preserving boundary conditions (SNP) and is defined as:
\ba
\Big \{ {\cal T}_1(\lambda),\  {\cal T}_2(\mu) \Big \} &=& r_{12}(\lambda-\mu)\ {\cal T}_1(\lambda)\ {\cal T}_2(\mu) - {\cal T}_1(\lambda)\ {\cal T}_2(\mu)\ r_{21}(\lambda-\mu) \cr &+& {\cal T}_1(\lambda)\ r_{21}^{t_{1}}(-\lambda-\mu)\ {\cal T}_2(\mu)\ - {\cal T}_2(\mu)\ r_{12}^{t_2}(-\lambda-\mu)\ {\cal T}_1(\lambda) \label{ty}
\ea
The generic representation of the classical reflection algebras is given by \cite{sklyanin, doikou-twist}
\be
{\cal T}_a(\lambda) =  T_a(L ,0; \lambda)\ K_a(\lambda)\ T_a^{t_a}(L, 0; -\lambda), \label{rep2}
\ee
recall $T_a(L ,0; \lambda) = T^+(L,x_0; \lambda)\ {\mathbb L}(x_0;\lambda)\ T^-(x_0, 0; \lambda)$, $K$ is a non-dynamical solution of the reflection equation (\ref{nondyn}), i.e.
\be
\{K_a(\lambda),\ K_b(\mu)\} =0. \label{nondyn}
\ee
The local integrals of motion are obtained via the generating function
\be
{\cal G}(\lambda) = \ln tr_a(\bar K_a(\lambda){\cal T}_a(\lambda)), \label{gf2}
\ee
where $\bar K$ is also a c-number (non-dynamical) (\ref{nondyn}) solution of the classical twisted Yangian.
We shall assume here for simplicity, but without loss of generality as will be evident later, Schwartz boundary condition at $x=0$, and $K \propto {\mathbb I}$. Also consider at $x=L$ the following $\bar K$ matrix \cite{gand, avan-doikou-toda}
\ba
&& \bar K(\lambda)  = u^{3\over 2}\bar G + u^{1\over 2} \bar F + u^{-{1\over 2}} F + u^{-{3\over 2}} G ~~~~\mbox{where} \cr && G = g\ {\mathbb I}, ~~~~~\bar G = \bar g\ {\mathbb G} \cr
&& \bar F = f_{12}\ e_{21} + f_{23}\ e_{32} + f_{31}\ e_{13} \cr
&& F = f_{21}\ e_{12} + f_{23}\ e_{32} + f_{13}\ e_{31} \cr
&& g = q^{1\over 4}, ~~~~~\bar g = \pm q^{-{1\over 4}}, ~~~~f_{ij}= \pm q^{-{3\over 4}}~~~~f_{ji}= q^{3\over 4} , ~~~i<j.
\ea

Expansion of the generating function (\ref{gf2}) in powers of $u^{-1}$ provides the following expressions:
\ba
{\cal G}(u) &=& Z_{33}^{+}(u) +\hat Z_{33}^+(u^{-1}) + Z_{33}^{-}(u) +\hat Z_{33}^-(u^{-1}) \cr
&+& \ln \Big [ (1 + \hat W^+(u^{-1};0))^t(\Omega^+(0))^{-1}\bar K(\lambda) \Omega^+(0) (1+W^+(u;0)) \Big ]_{33}\cr
&+& \ln \Big [(1+ W^+(u;x_0))^{-1} (\Omega^+(x_0))^{-1} {\mathbb L}(u; x_0) \Omega^-(x_0) (1+W^-(u;x_0))\Big ]_{33} \cr
&+& \ln \Big [(1+ \hat W^-(u^{-1};x_0))^t (\Omega^-(x_0))^{-1} {\mathbb L}^t(x_0; u^{-1}) \Omega^+(x_0) ((1+\hat W^+(u^{-1};x_0))^{-1})^t\Big ]_{33}
\ea
The boundary contribution at $x  = - L $ is trivial. Due to the fact that for any matrix $A$  $A_{jj} = A^t_{jj}$ it is clear that the last term of the latter expression essentially coincides with the defect contribution in (3.22) ($u \to u^{-1}$), thus in this case the first non-trivial integral of motion is in fact the Hamiltonian expressed as
\be
H = - {6 m \over \beta^2} \Big ( Z_{33}^{+(1)} + \hat Z_{33}^{+(1)} + Z_{33}^{-(1)}  + \hat Z_{33}^{-(1)} + (X^{(0)})^{-1} X^{(1)} + (\hat X^{(0)})^{-1} \hat X^{(1)}\Big ) + \mbox{boundary term}
\ee
Note that explicit computations on the boundary terms were performed in \cite{avan-doikou-toda}. In this case, as expected due to the presence of the non-trivial boundaries the momentum is not a conserved quantity anymore (see also e.g.\cite{corrigan-twist, doikou-twist}). The explicit expression of the associated Hamiltonian is then given by:
\ba
&& H = \int_{0}^{x_0^-} dx \sum_{i=1}^2 \Big( (\pi_i^-)^2 + (\phi_i^{-'})^{2}
 + {m^2 \over \beta^2} \sum_{i=1}^3 e^{\beta \alpha_i \cdot \Phi^-}\Big) \cr
 && + \int_{x_0^+}^{L} dx \sum_{i=1}^2 \Big( (\pi_i^+)^2+(\phi_i^{+'})^{2} +{m^2 \over \beta^2} \sum_{i=1}^3 e^{\beta \alpha_i \cdot \Phi^+}\Big) + \sum_{i=1}^3 c_i e^{\beta \alpha_i\cdot \Phi^{+}(L)}\cr
&& -{6m \over \beta^2} \Big ({4 \over 3m} (c^- + b^+ -b^- - c^+-\hat \alpha^- + \hat c^-+\hat \alpha^+- \hat c^+) + {1 \over 3} (T_{13} - T_{21} - T_{32}-\hat T_{31} + \hat T_{12} + \hat T_{23})\Big )\Big |_{x=x_0}. \cr &&
\ea
The last term in the second line of the latter expression is the boundary term coinciding as expected with the terms extracted in \cite{corrigan-twist, avan-doikou-toda} (see \cite{avan-doikou-toda} for a more detailed description on the boundary contributions). It is clear that the defect contribution in the Hamiltonian is not altered compared to the periodic case studied in the previous section (see also for analogous findings in \cite{nls-twist} in the vector non-linear Schr\"{o}dinger context).

\subsection*{The Lax pair}

Explicit expressions of the time components of the Lax pairs in the bulk as well as on the defect point when integrable boundaries conditions are present are provided in \cite{nls-twist}. Let us first focus on the bulk expressions; in \cite{nls-twist} the fundamental Poisson bracket was formulated:
\be
\Big \{\bar K_a(\lambda){\cal T}_a(\lambda),\ {\mathbb U}^{\pm}_b(\mu) \Big \} = {\partial \over \partial x}\Big ({\cal M}^{\pm}_{ab}(\lambda,\mu) + {\cal M}_{ab}^{\pm *}(\lambda,\mu) \Big ) + \Big [ {\cal M}^{\pm}_{ab}(\lambda, \mu)+ {\cal M}_{ab}^{\pm*}(\lambda, \mu) ,\ {\mathbb U}_{b}(\mu)\Big ]
\ee
and ${\cal M}^{\pm},\ {\cal M}^{\pm *}$ are defined as:
\ba
&& {\cal M}_{ab}^{\pm}(\lambda, \mu) = \bar K_a(\lambda)\ M_{ab}^{\pm}(\lambda, \mu)\ K_a(\lambda)\ \ T_a^{t_a}(-\lambda) \cr
&& {\cal M}_{ab}^{\pm *}(\lambda, \mu) = \bar K_a(\lambda)\ T_a(\lambda)\  K_a(\lambda)\ M_{ab}^{\pm*}(\lambda,\mu).
\ea
$M^{\pm}$ are defined in (\ref{mm}) (but $-L \to 0$), and
\be
M^{\pm *}_{ab}(\lambda, \mu) = \Big ( M_{ab}^{\pm}(-\lambda,\mu)\Big )^{t_a}.
\ee
After considering the generating function of the local integrals of motion ${\cal G}(\lambda)$ we conclude that:
\be
\dot {\mathbb U}^{\pm}(\mu) = {\partial {\mathbb V}^{\pm}(\lambda, \mu) \over \partial x } +
\Big [ {\mathbb V}^{\pm}(\lambda, \mu),\ {\mathbb U}^{\pm}(\mu) \Big ],
\ee
where the bulk quantities ${\mathbb V}^{\pm}$ are expressed as (they are defined up to an overall normalization factor):
\be
{\mathbb V}_b^{\pm}(\lambda, \mu; x) = t^{-1}(\lambda)\ tr_a \Big ({\cal M}_{ab}^{\pm}(\lambda, \mu) +
{\cal M}_{ab}^{\pm *}(\lambda, \mu) \Big ) ~~~~~x \neq x_0. \label{VV}
\ee
Expansion of ${\mathbb V}^{\pm}$ in powers of $u^{-1}$ will provide the time components of the Lax pairs associated to each local integral of motion.

Special care is taken on the defect point. Taking into account the zero curvature condition on the defect point (\ref{zcd}), as well as certain algebraic relations the following Poisson bracket was formulated in \cite{nls-twist}:
\be
\Big \{\bar K_a(\lambda)\ {\cal T}_a(\lambda),\ {\mathbb L}_b(\mu) \Big \} =\Big ( \tilde {\cal M}^{+ }_{ab}(\lambda,\mu)+ \tilde {\cal M}_{ab}^{+*}(\lambda, \mu)\Big )\ {\mathbb L}_b(\mu) - {\mathbb L}_b(\mu)
\ \Big (\tilde {\cal M}^{-}_{ab}(\lambda, \mu) + \tilde {\cal M}_{ab}^{-*}(\lambda, \mu)\Big)
\ee
where
\ba
&& \tilde {\cal M}_{ab}^{\pm}(\lambda, \mu) = \bar K_a(\lambda)\ \tilde M^{\pm}_{ab}(\lambda, \mu)\ K_a(\lambda)\ T_a^{t_a}(-\lambda)  \cr
&&  \tilde {\cal M}_{ab}^{\pm*}(\lambda, \mu) = \bar K_a(\lambda)\ T_a(\lambda)\  K_a(\lambda)\ \tilde M^{\pm *}_{ab}(\lambda, \mu),
\ea
and we define
\be
\tilde M^{\pm*}_{ab}(\lambda, \mu) = \Big ( \tilde M_{ab}^{\pm}(-\lambda,\mu)\Big )^{t_a}.
\ee

The aim now is to identify $\tilde {\mathbb V}^{\pm}(x_0)$, i.e. the Lax pairs on the defect point from the left and from the right. Indeed, bearing in mind the zero curvature condition on the defect point we may directly identify the relevant $\tilde {\mathbb V}^{\pm}$ operators (defined up to an overall normalization factor)
\be
\tilde {\mathbb V}_b^{\pm}(\lambda, \mu; x_0) = t^{-1}(\lambda)\ tr_a \Big ( \tilde {\cal M}^{\pm }_{ab}(\lambda, \mu) + \tilde {\cal M}^{\pm *}_{ab}(\lambda, \mu)\Big ). \label{tvv}
\ee

Expansion in powers of $u^{-1}$ provide the relevant Lax pair. For instance the first order in the expansion is $\tilde {\mathbb V}_H$ derived in (\ref{vv}). Then analyticity conditions imposed on the time components of the Lax pairs
\be
\tilde {\mathbb V}^{\pm(n)}(x_0^{\pm}) \to {\mathbb V}^{\pm(n)}(x_0) \label{cont}
\ee
will provide the wanted sewing conditions as explained in \cite{avan-doikou1}.
In this case only ``half'' of the sewing constraints arise due to the reduced number of concerned quantities
\ba
&& \phi_1^{+'} - \phi_1^{-'}= {m\over 4 \beta} \Big ( 2T_{21} -T_{32} +T_{13} -2\hat T_{12} + \hat T_{23} -\hat T_{31} \Big ) \cr
&& \phi_2^{+'} - \phi_2^{-'} = {\sqrt{3} m \over 4 \beta} \Big (T_{32} + T_{13} -\hat T_{23} - \hat T_{31} \Big ).
\ea
With this we complete our analysis on the $A_2^{(1)}$ ATFT in the presence of local defects and ``twisted'' boundary conditions.
The boundary conditions in ATFT's have been studied in detail in \cite{corrigan-twist, avan-doikou-toda}, what happens at the boundary point $x=0$ in particular is fully described there, therefore we do not repeat the analysis here. However, we refer the interested reader to \cite{corrigan-twist, avan-doikou-toda} for the detailed analysis on the boundary conditions in ATFT's.

\section{Discussion}

Let us summarize the findings of the present study. We have considered here the $A_{2}^{(1)}$ ATFT in the presence of a point-like integrable defect, and have employed the Hamiltonian formalism for the analysis of the problem. The model is examined in the case of Schwartz boundary conditions as well as in the presence of ``twisted'' boundary conditions. In the case where boundaries are implemented half of the charges as expected are conserved e.g. the Hamiltonian is still a conserved quantity, whereas the momentum is not conserved anymore. Nevertheless, these particular boundary conditions do not alter the defect contribution in the Hamiltonian expression a well as in the respective Lax pair as was also observed in the vector non-linear Schr\"{o}dinger model \cite{nls-twist}.

Boundary conditions associated to the reflection algebra are of particular interest given that they may drastically alter the defect behavior as was already demonstrated in the vector non-linear Schr\"{o}dinger model in \cite{nls-twist}. This issue due to its own significance, but also due to various technical challenges will be separately addressed in a forthcoming publication. Moreover, identification of the classical scattering between solitonic solutions and the defect via the inverse scattering method and comparison with the corresponding quantum results (see e.g. \cite{fus, konle, BCZ3, doikou-karaiskos, doikou-defects}) would be of particular significance. Solitonic solutions in this context can be derived through the associated B\"{a}cklund transformations. This is an interesting issue, in particular regarding the defect point, and its connection to the integrable sewing conditions.

\appendix

\section{Identification of the $W,\ Z\ (\hat W,\ \hat Z)$ matrices}

We identify here the first order contributions on expressions (\ref{wz}) using (\ref{dec}), (\ref{dec2}) and (\ref{dif1}).
More precisely, substitution of the ansatz (\ref{dec}), (\ref{dec2}) in (\ref{dif1}) gives rise to Riccati type differential equations \cite{ft, avan-doikou-toda} for the non-diagonal part leading to:
\ba
&& W^{\pm(0)} = \hat W^{\pm(0)} =
\begin{pmatrix}
0& e^{{i\pi \over 3}} &1 \\
 e^{{i\pi \over 3}}& 0&-1\\
e^{{2i\pi \over 3}} &e^{-{i\pi \over 3}}   &0
\end{pmatrix} \cr
&& {m\over 4}  W^{\pm(1)} =
\begin{pmatrix}
0& e^{{2i\pi \over 3}}a^{\pm} &c^{\pm} \\
-a^{\pm}& 0&b^{\pm}\\
e^{{i\pi \over 3}}c^{\pm} &-b^{\pm}   &0
\end{pmatrix}, ~~~~~
{m\over 4} \hat W^{\pm(1)} =
\begin{pmatrix}
0& -\hat b^{\pm}& -\hat a^{\pm} \\
-e^{-{i\pi \over 3}}\hat b^{\pm} & 0&-\hat c^{\pm}\\
\hat a^{\pm} &-e^{{i\pi \over 3}}\hat c^{\pm}   &0
\end{pmatrix}
\ea

The diagonal elements are also given below.
The behavior of $Z^{\pm(-1)}, (\hat Z^{\pm(-1)})$ is important, because this will determine the leading contribution in the transfer matrix expansion as $|u| \to \infty,\ {|u^{-1}| \to \infty}$
\be
Z^{\pm(-1)}(x,y) = {m(x-y) \over 4}
\begin{pmatrix}
e^{{i\pi\over 3}}& &  \\
 & e^{-{i\pi\over 3}}&\\
&   &-1
\end{pmatrix}, ~~~~~~\hat Z^{\pm(-1)}(x,y) = {m(x-y) \over 4}
\begin{pmatrix}
e^{-{i\pi\over 3}}& &  \\
 & e^{{i\pi\over 3}}&\\
&   &-1
\end{pmatrix}.
\ee
Also, the next order contributions are given by:
\be
{d Z^{\pm(0)} \over dx} = 0, ~~~~~\ {d \hat Z^{\pm(0)} \over dx} =0,
\ee
and
\ba
&& {d Z_{11}^{\pm(1)} \over dx} = {m e^{- {i\pi \over 3}}\over 12} (\gamma_1^{\pm} + \gamma_2^{\pm} + \gamma_3^{\pm}) + {4 e^{-{i\pi \over 3}} \over 3m} (a^{\pm'} -c^{\pm'} ) + {2 e^{-{i\pi \over 3}} \over 3m} (a^{\pm 2} + b^{\pm 2} +c^{\pm 2})\cr
&& {d Z_{22}^{\pm(1)} \over dx} = {m e^{{i\pi \over 3}} \over 12} (\gamma_1^{\pm} + \gamma_2^{\pm} + \gamma_3^{\pm} ) + {4 e^{{i\pi \over 3}} \over 3m} (b^{\pm'} -a^{\pm'})+ {2 e^{{i\pi \over 3}} \over 3m} (a^{\pm 2} + b^{\pm 2} +c^{\pm 2}) \cr
&& {d Z_{33}^{\pm(1)} \over dx} = -{m\over 12} (\gamma_1^{\pm} + \gamma_2^{\pm} + \gamma_3^{\pm}) - {4  \over 3m} (c^{\pm '} -b^{\pm '}) - {2 \over 3m} (a^{\pm 2} + b^{\pm 2} +c^{\pm 2})
\ea
\ba
&& {d \hat Z_{11}^{\pm(1)} \over dx} = {m e^{ {i\pi \over 3}} \over 4} (\gamma_1^{\pm} + \gamma_2^{\pm} + \gamma_3^{\pm}) - {4 e^{{i\pi \over 3}} \over 3m} (\hat b^{\pm'} - \hat a^{\pm'} ) + {2 e^{{i\pi \over 3}} \over 3m} (\hat a^{\pm 2} + \hat b^{\pm 2} +\hat c^{\pm 2})\cr
&& {d \hat Z_{22}^{\pm(1)} \over dx} = {m e^{- {i\pi \over 3}} \over 4} (\gamma_1^{\pm} + \gamma_2^{\pm} + \gamma_3^{\pm} ) + {4 e^{-{i\pi \over 3}} \over 3m} (\hat b^{\pm'} -\hat c^{\pm'})+ {2 e^{-{i\pi \over 3}} \over 3m} (\hat a^{\pm 2} + \hat b^{\pm 2} +\hat c^{\pm 2}) \cr
&& {d \hat Z_{33}^{\pm(1)} \over dx} = -{m\over 12} (\gamma_1^{\pm} + \gamma_2^{\pm} + \gamma_3^{\pm}) + {4  \over 3m} (\hat a^{\pm '} -\hat c^{\pm '}) - {2 \over 3m} (\hat a^{\pm 2} + \hat b^{\pm 2} +\hat c^{\pm 2})
\ea
where we define:
\ba
&& a^{\pm} = {\beta \over 2} \Big ( {\theta_1^{\pm} \over 2} + {\theta_2^{\pm} \over 2{\sqrt 3}}\Big ), ~~~~b^{\pm} = {\beta \over 2} \Big ( -{\theta_1^{\pm} \over 2} + {\theta_2^{\pm} \over 2{\sqrt 3}}\Big ), ~~~~c^{\pm} = -{\beta \over 2} {\theta_2^{\pm} \over {\sqrt 3}}, ~~~\theta_i = \pi_i - \phi'_i \cr
&& \gamma_1^{\pm} = e^{\beta \phi_1^{\pm}}, ~~~~\gamma_2^{\pm} = e^{{\beta\over 2}(-\phi_1^{\pm}+\sqrt{3} \phi_2^{\pm})}, ~~~~\gamma_3^{\pm} = e^{-{\beta\over 2}(\phi_1^{\pm}+\sqrt{3} \phi_2^{\pm})}.
\ea
Similarly $\hat a^{\pm}, \hat b^{\pm}, \hat c^{\pm}$ are defined as $a^{\pm},\ b^{\pm},\ c^{\pm}$, but with $\theta^{\pm}_i \to \hat \theta_i$, $\hat \theta^{\pm}_i = \pi^{\pm}_i + \phi^{\pm'}_i$.

\end{document}